\documentclass[aps,prb,showpacs,twocolumn,amsmath,amssymb,footinbib]{revtex4}
\usepackage{latexsym}
\usepackage{graphicx,epstopdf}
\usepackage{subfigure}
\usepackage{epsfig}
\usepackage{amsfonts}
\usepackage{amssymb}
\usepackage{amsmath}
\usepackage{bm}
\usepackage{braket}
\usepackage{color}
\usepackage{mathdots}

\newcommand{\tr}{{\rm tr}\;}
\renewcommand{\i}{{\rm i}}
\newcommand{\e}{{\rm e}}
\begin{document}

\title{One-dimensional lattice model with an exact matrix-product ground
state\\ describing the Laughlin wave function}

\author{Zheng-Yuan Wang and Masaaki Nakamura\cite{address}}
\affiliation{Department of Physics, Tokyo Institute of Technology, O-Okayama,
Meguro-ku, Tokyo 152-8551, Japan}

\date{\today}

\begin{abstract}
 We introduce one-dimensional lattice models with exact matrix-product
 ground states describing the fractional quantum Hall (FQH) states in
 Laughlin series (given by filling factors $\nu=1/q$) on torus geometry.
 Surprisingly, the exactly solvable Hamiltonian has the same
 mathematical structure as that of the pseudopotential for the Laughlin
 wave function, and naturally derives the general properties of the
 Laughlin wave function such as the $Z_2$ properties of the FQH states
 and the fermion-boson relation.
 The obtained exact ground states have high overlaps with the Laughlin
 states and well describe their properties.  Using the matrix product
 method, density functions and correlation functions are calculated
 analytically. Especially, obtained entanglement spectra reflects
 gapless edge states as was discussed by Li and Haldane.
\end{abstract}
\pacs{71.10.Pm, 75.10.Kt, 73.43.Cd}

\maketitle 

\section{Introduction}
The fractional quantum Hall (FQH) effect has been one of the most
intriguing research fields in condensed-matter physics during the past
few decades, in both theory and experiment.\cite{QH,Laughlin}  Recently
the FQH effect has attracted much attentions due to new developments
including the observation in graphene,\cite{fqhgraphene} flat band
Chern insulators \cite{chernins}, rapidly rotating bosons,\cite{boseQH}
applications to topological quantum computing,\cite{qcomp} and so on.


A key property of FQH states is their topological order. One consequence
thereof is that their physical properties are insensitive to smooth
deformations of the manifold on which we choose to study them.  Using
toroidal boundary conditions, the two-dimensional (2D) continuum system
in a magnetic field can be reduced to a one-dimensional (1D) lattice
model.  This fact has been exploited in a series of recent studies of
the interacting many-body problem in the limit geometry of a thin torus,
referred as the Tao-Thouless (TT)
limit.\cite{Tao-T,Rezayi-H,Bergholtz-K,Seidel-F-L-L-M}


In this paper, based on the 1D approach, we introduce minimal models
with exact ground states that describe the FQH states in the Laughlin
series (filling factors $\nu=1/q$).  The physical quantities,
such as density functions and correlation functions,
are calculated analytically by the matrix product (MP) method.
The simplest case of our model
($q=3$) reproduces a model describing the $\nu=1/3$ FQH state that has
been discussed quite recently by the authors with Bergholtz,
\cite{Nakamura-W-B} and Jansen.\cite{Jansen}
We will also show that the structure of the 1D models
[Eq.~(\ref{Ham1q})-(\ref{opq-b})below]
implies that of the pseudopotential for the Laughlin wave function [Eq.~(\ref{TKpot}) below],
so that properties of the Laughlin wave function such as
the $Z_2$ classification (the FQH effect is limited only odd $q$ for fermions
and even $q$ for bosons) and general
relations between fermionic and bosonic systems are naturally derived
from our model.

Among physical quantities obtained by the MP method, entanglement
spectra (ES) have attracted the attention recently.  ES are obtained as
coefficients when the wave function is divided into two subsystems.
According to Li and Haldane.\cite{Li-Haldane}, ES describe the ``edge
states'' of the subsystem.  Based on this notion, we extract (chiral)
Tomonaga-Luttinger liquid behavior of the edge state analytically.

This paper is organized as follows: In Sec.~\ref{sec:model}, we discuss
how 1D models with exact ground states are related to the $\nu=1/q$
Laughlin states. We also consider bosonic version of the model.  In
Sec.~\ref{sec:MP}, we introduce MP representation of the exact ground
state, and perform analytical calculation of the physical quantities
such as density functions, correlation functions, and ES.
In Appendices,
we present the detailed calculation of the matrix elements of the pseudopotential,
and those of the physical quantities based on the MP method.

\section{Models with exact ground states}\label{sec:model}

\subsection{FQH system on a torus}
As discussed in the preceeding
works,\cite{Bergholtz-K,Seidel-F-L-L-M,Nakamura-W-B,Jansen} when we
consider 2D fermion system in a magnetic field using the torus boundary
condition with lengths $L_l$ in $x_l$ directions $(l=1,\,2)$, and the
single-particle wave function with the Landau gauge $\bm{A}=Bx_2\hat{x}_1$ for
the lowest Landau level, the system is described by the following model
on a 1D discrete lattice on direction $x_2$:
\begin{equation}
\mathcal{H}
=\sum_{i=1}^{N_s}
\sum_{|m|<k\leq N_s/2}
 V_{km}
 c_{i+m}^{\dag}
 c_{i+k}^{\dag}
 c_{i+m+k}^{\mathstrut}
 c_{i}^{\mathstrut},
 \label{1D_model}
\end{equation}
where $c_i^\dagger$ ($c_i$) creates (annihilates) a fermion at site $i$,
$\hat n_i\equiv c_i^\dagger c_i^{\mathstrut}$, and $N_s=L_1 L_2/2\pi$ is
the number of sites (the magnetic length $l_{\rm B}\equiv
\sqrt{\hbar/eB}$ is set to be unity).  The matrix element $V_{km}$
specifies the amplitude for a process where particles with separation
$k+m$ hop $m$ steps to a separation $k-m$.  This model conserves the
center-of-mass coordinate defined by $K=\sum_{j=1}^{N_s} j \hat{n}_j \
({\rm mod} \ N_s)$, which is related to the momentum of the $x_1$ direction.
\subsection{$\nu=1/3$}
In the preceding paper,\cite{Nakamura-W-B,Jansen} a 1D model with an
exact ground state describing the $\nu=1/3$ FQH state is constructed in
the following way: First, as an approximation in the vicinity of the TT
limit, we truncate the Hamiltonian (\ref{1D_model}) at $\nu=1/3$ up to
the four leading terms ($k+|m|\leq 3$). In the condition
\begin{equation}
 V^2_{21}=V_{10}V_{30},\qquad V_{20}>0,
  \label{condition13}
\end{equation}
which is satisfied for the $\nu=1/3$ Laughlin state for any $L_1$ with
$L_2\to\infty$, this model can be rewritten as
\begin{equation}
\mathcal{H}_{1/3}=
 \sum_{i} [Q^\dagger_i Q_i^{\mathstrut} +
 P^{\dagger}_i P_i^{\mathstrut}],
\end{equation}
where 
\begin{equation}
Q_i=\alpha_1 c_{i+1}c_{i+2}+\alpha_0 c_{i}c_{i+3},\quad
P_i=\beta_0 c_{i}c_{i+2}
\end{equation}
with $\alpha_0, \alpha_1, \beta_0 \in {\mathbf R}$.  This Hamiltonian
apparently has positive expectation values
$\braket{\mathcal{H}_{1/3}}\geq 0$.  Next, we introduce the wave
function
\begin{equation}
 \ket{\Psi_{1/3}}=\prod_i(\alpha_1 - \alpha_0 c^\dagger_{i+1}
  c^\dagger_{i+2} c_{i+3}^{\mathstrut}
  c_{i}^{\mathstrut})\ket{\Psi^0_{1/3}}, 
\end{equation}
with $\ket{\Psi^0_{1/3}}\equiv\ket{100100100\cdots}$. Then the relation
$Q_i\ket{\Psi_{1/3}}=P_i\ket{\Psi_{1/3}}=0, \forall i$ is satisfied,
so that $\ket{\Psi_{1/3}}$ is shown to be an exact ground state of
$\mathcal{H}_{1/3}$. This state is three-fold degenerate due to the
center-of-mass conservation. This argument can also be applied to a
$\nu=1/2$ bosonic FQH state with simple replacements.

\subsection{$\nu=1/q$}
Now we discuss how an exactly solvable model is constructed for general
Laughlin series $\nu=1/q$ with odd $q$.
For $\nu=1/q$ with $q>3$, the Hamiltonian should include longer range
interactions than $\mathcal{H}_{1/3}$.  The dimension of the subspace
increases as $(q+1)/2=[q/2]+1$ where $[\cdots]$ is the integer part of
$\cdots$ (e.g.  $\left\{\ket{1001}, \ket{0110}\right\}$ for $q=3$ and
$\left\{\ket{100001}, \ket{010010}, \ket{001100}\right\}$ for $q=5$).
We find that the extended model should be given in the following form:
\begin{subequations}
 \begin{align}
 \mathcal{H}_{1/q}&=\sum_{l=1}^{[q/2]}
 \sum_{i} [Q^{(l)\dagger}_i Q_i^{(l)\mathstrut}
   + P^{(l)\dagger}_i P_i^{(l)\mathstrut}],
   \label{Ham1q}\\
  Q_{i}^{(l)}&\equiv\sum_{\mu=0}^{[q/2]}
  \alpha_{\mu}^{(l)} c_{i+\mu} c_{i+q-\mu},\label{opq-a}\\
  P_{i}^{(l)}&\equiv\sum_{\mu=0}^{[q/2]-1}
  \beta_{\mu}^{(l)} c_{i+\mu} c_{i+q-\mu-1}.\label{opq-b}
 \end{align}
   \label{Ham1q_all}
\end{subequations}
This is also positive semidefinite $\braket{\mathcal{H}_{1/q}}\geq 0$.
We can construct a similar Hamiltonian with zero energy eigenstates if
we take only single-$l$ term in (\ref{Ham1q}) into account as was done
in Refs.~\onlinecite{Seidel-L}~and~\onlinecite{Qi}.
 However, if there are no degrees of freedoms of $l$,
we cannot fix the parameters $\alpha_{\mu}^{(l)}$ uniquely,
as will be discussed below.

The corresponding wave function which gives the zero energy eigenstate is
\begin{equation}
 \ket{\Psi_{1/q}}=\prod_i\sum_{j=0}^{[q/2]}
  s_{j}\hat{U}_{q-j,j}(i) \ket{\Psi_{1/q}^0},
  \label{gwf1q}
\end{equation}
where the squeezing operators are defined by $\hat{U}_{k,0}\equiv1$ and
$\hat{U}_{k,m}(i)\equiv c_{i+m}^{\dag}
c_{i+k}^{\dag}c_{i+k+m}^{\mathstrut} c_i^{\mathstrut}$ for $m\neq0$.
This operator is identical with a part of the pair hopping terms in
Eq.~(\ref{1D_model}).  The root wave function $\ket{\Psi_{1/q}^0}$ is
given by the charge-density-wave state in the TT limit,
\begin{equation}
 \ket{\Psi_{1/q}^0}=
  \ket{\cdots\underbrace{10\cdots0}_{q}
  \underbrace{10\cdots0}_{q}\cdots},
\end{equation}
where every $q$th site is occupied.  The parameters in
Eqs.~(\ref{opq-a}),~(\ref{opq-b})~and~(\ref{gwf1q}) are determined so that they satisfy
orthogonal conditions, $\bm{\alpha}^{(l)}\cdot\bm{s}=0, \forall l$ where
$\bm{\alpha}^{(l)}=(\alpha^{(l)}_0,\alpha^{(l)}_1,
\cdots,\alpha^{(l)}_{[q/2]})$ and $\bm{s}=(1,s_1,\cdots,s_{[q/2]})$.
{\it Since the number of the subspace (components of $\bm{s}$) is larger
than the number that $l$ may take by 1, these equations fix the
parameters $\bm{s}$ uniquely.}  On the other hand, there is no
constraint for the parameter $\beta_{\mu}^{(l)}$, since the operator
$P_{i}^{(l)}$ gives pair hopping processes between electrons separated even
number of lattice sites which do not exist in Eq.~(\ref{gwf1q}). Thus
$Q_i^{(l)}\ket{\Psi_{1/q}}=P_i^{(l)}\ket{\Psi_{1/q}}=0, \forall i$ is
satisfied which means that (\ref{gwf1q}) is the exact ground state of
Eq.~(\ref{Ham1q}) with zero energy. Due to the center-of-mass
conservation, the ground state is $q$-fold degenerate,
even if parameters $\bm{\alpha}$, $\bm{\beta}$, and $\bm{s}$ have site dependence.
Uniqueness of the ground state in each center-of-mass sector
can be shown considering that other states have finite positive energies.

\subsection{Bosonic systems}
The similar argument can also be applied to bosonic FQH systems with
even $q$.  The bosonic version of (\ref{1D_model}) is given by
\begin{equation}
 \mathcal{H}
  =\sum_{i=1}^{N_s}
  \sum_{|m|\leq k\leq N_s/2}
  V_{km}
  b_{i+m}^{\dag}
  b_{i+k}^{\dag}
  b_{i+m+k}^{\mathstrut}
  b_{i}^{\mathstrut},
 \label{1D_model_b}
\end{equation}
where $b_i^\dagger$ ($b_i$) creates (annihilates) a boson at site $i$.
In this case the dimension of the subspace is $(q+2)/2$ for even $q$
(e.g. $\left\{\ket{10001}, \ket{01010}, \ket{00200}\right\}$ for $q=4$).
Our exact argument for fermions can be applied straightforwardly to the
bosonic cases only replacing the fermionic operators by the bosonic ones
$c_i^{(\dag)}\to b_i^{(\dag)}$ in Eqs.~(\ref{opq-a}),~(\ref{opq-b})~and~(\ref{gwf1q}).

\subsection{Relationship with Laughlin states}
We discuss that our exactly solvable model well describes the $\nu=1/q$
Laughlin states.  In order to discuss the Laughlin sates in the present
1D framework, we introduce a pseudopotential,\cite{Haldane,Trugman-K}
which is known to have the $\nu=1/q$ Laughlin state as the exact ground
state,
\begin{equation}
 \mathcal{V}(\bm{x})=\sum_{\lambda=0}^{q-1}
  c_{\lambda}b^{2\lambda}
  \nabla^{2\lambda}\delta^2(\bm{x}),
 \label{TKpot}
\end{equation}
where $\bm{x}\equiv(x_1,x_2)$, $c_{\lambda}=(-1)^{\lambda}|c_{\lambda}|$
are constants to keep the energy positive, and $b>0$ stands for the
range of the interactions.  Note that number of the terms increases as
$q$ is increased. This reminds us that our exactly solvable model
(\ref{Ham1q}) has a similar structure.  For $N_s \gg L_1$, the
single-particle wave function for the lowest Landau level in Landau
gauge $\bm{A}=Bx_2\hat{x}_1$ is given by
\begin{equation}
 \psi_k(\bm{x})\propto
  \mathrm{e}^{\i (2\pi k/L_1)x_1 -\frac{1}{2}(x_2+ 2\pi k/L_1)^2},
  \label{1particle_wf}
\end{equation}
where $2\pi k/L_1$ is the momentum along the $x_1$-direction.  In this
basis, we derive the general form of the matrix elements of the
Hamiltonian (\ref{1D_model}) as the second quantization of
(\ref{TKpot}),
\begin{equation}
\begin{split}
 V_{km}'=&\sum_{\lambda=0}^{q-1} V_{km}^{\prime(\lambda)}; \\
 V_{km}^{\prime(\lambda)}=&
\sum_{\nu=0}^{\lambda}
\sum_{\mu=0}^{[(\lambda-\nu)/2]}
 C_{\mu\nu}
 (k^2 -m^2)^{\nu}\\
 &\times\left[ (-1)^{\nu}k^{2\mu}\mp m^{2\mu}\right] 
 \mathrm{e}^{- \frac{2\pi^2}{L_1^2} (k^2+m^2)},
\end{split}
\label{TKint}
\end{equation}
where $C_{\mu\nu}$ are given by $c_{\lambda}$, $b$, and $L_1$ (see
Appendix \ref{sec:TKpotC}).  Note that the function is antisymmetric
(symmetric) under the exchange of $k^2$ and $m^2$ as a consequence of
the Fermi (Bose) statistics.

On the other hand, the matrix elements of our exactly solvable model
$V_{km}$ $(k+|m| \le q)$ are identified in the following way. First, we
divide the matrix elements as $V_{km}=\sum_{\lambda}V_{km}^{(\lambda)}$
and relate $\lambda$ to the variable in Eq.~(\ref{Ham1q}) as
$\lambda=2l-1$ for fermions $\lambda=2l$ for bosons. The matrix elements
for each $\lambda$ which stem from $[\cdots]$ in Eq.~(\ref{Ham1q})
satisfy the relation
\begin{equation}
 \left[V^{(\lambda)}_{km}\right]^2
  =V^{(\lambda)}_{k-m,0}V^{(\lambda)}_{k+m,0}. 
  \label{condition1q}
\end{equation}
This relation is a counterpart of Eq.~(\ref{condition13}) for the
$\nu=1/q$ case, and is also satisfied if we choose the matrix elements
in the following form:
\begin{equation}
\begin{split}
 V_{km}=&\sum_{\scriptstyle \lambda=1, {\rm odd}
 \atop \scriptstyle \lambda=0, {\rm even}}^{q-1}
 V_{km}^{(\lambda)},\\
 V_{km}^{(\lambda)}
 =&C_{\lambda}(k^2 -m^2)^{\lambda}
 \mathrm{e}^{-\frac{2 \pi^2}{L_1^2}(k^2+m^2)},
\end{split} 
 \label{effTK}
\end{equation}
where $C_{\lambda}$ is a constant and $\lambda$ is restricted to be odd
(even) for fermions (bosons) due to the statistics, as already
mentioned.  Although Eqs.~(\ref{TKint}) and (\ref{effTK}) coincide at
$q=3$ ($q=2$) for fermions (bosons), they are different in general.
However, we may interpret the truncated Hamiltonian of (\ref{1D_model})
with (\ref{effTK}) as an approximation of the full Hamiltonian with
(\ref{TKint}) for $L_1\to 0$ limit (see Appendix \ref{sec:TKpotC}).

\begin{figure}[t]
 \includegraphics[width=80mm]{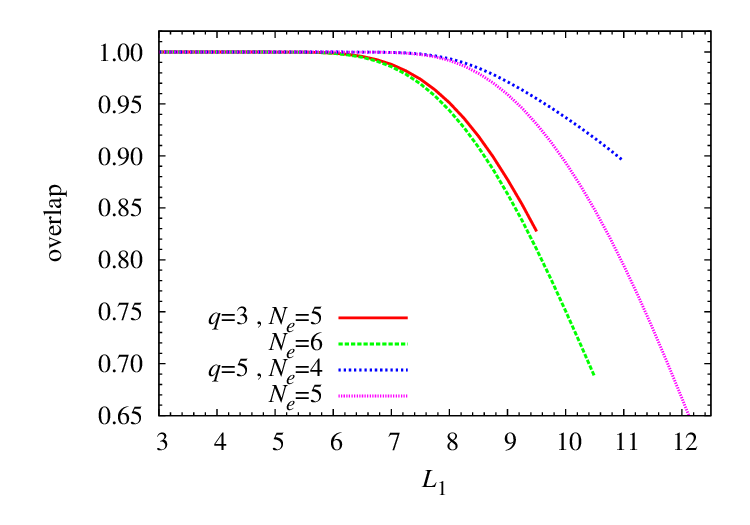}
 \caption{(Color online) Overlap of the ground-state wave function of
 the fermionic Laughlin state and that of our exactly solvable model
 (\ref{gwf1q}) for $q=3,5$ and $N_e\equiv N_s/q$ as functions of $L_1$
 in finite-size systems, calculated by exact diagonalization. The data
 are plotted up to the isotropic points of the torus ($L_1=L_2$). The
 approximation becomes better for larger-$q$ cases, since the number of
 long-range terms are added as $q$ is increased.} \label{fig1}
\end{figure}

In order to show that our model well describes the Laughlin state, we
numerically calculate overlap of the ground-state wave function of the
Laughlin state and that of our exactly solvable model (\ref{gwf1q}) in
finite-size systems using exact diagonalization.  As shown in
Fig.~\ref{fig1}, the overlap stays close to 1 up to $L_1\sim 6$ for all
cases, and the high overlap regions are extended for larger $q$.  Thus
we may conclude that our wave function well describes the Laughlin
states at $\nu=1/q$ with $q\geq 3$ ($q\geq 2$) for fermions (bosons).
Note that the approximation becomes better for larger-$q$ cases, since
the number of long-range terms are added as $q$ is increased. This is in
contrast with the Affleck-Kennedy-Lieb-Tasaki (AKLT)\cite{AKLT} model as
an approximation of the integer-spin Heisenberg model
which includes more extra terms as the length of the spin is increased.

\subsection{$Z_2$ properties}
The present exact result seems to be applicable to fillings $\nu=1/q$
with even (odd) denominator $q$ for fermions (bosons) as well.
However, the structure of our model naturally excludes such a situation,
in accordance with the symmetry of the Laughlin wave function.\cite{Laughlin}
For fermions (bosons) with even (odd) $q$,
the expressions of $Q_{i}^{(l)}$, $P_{i}^{(l)}$, and Eq.~(\ref{effTK}) remain unchanged,
while the summations in Eqs.~(\ref{Ham1q}) and (\ref{gwf1q}) are replaced by
$\sum_{l=1}^{[(q+1)/2]}$ and $\sum_{j=0}^{[(q-1)/2]}$, respectively.
This is because the number of subspace for fermions (bosons) with
even (odd) $q$ system becomes $[(q+1)/2]$ (e.g., $\left\{\ket{10001},
\ket{01010}\right\}$ for $q=4$ fermions).
Thus the number that $l$ may take becomes the same as the
number of the subspace.  Therefore, the parameter $\bm{s}$ cannot be
determined uniquely for these cases by the equation
$\bm{\alpha}^{(l)}\cdot\bm{s}=0, \forall l$, which means that models
with the exact-ground states (\ref{gwf1q}) cannot be constructed.

\subsection{Compressibility}
We can show the vanishing of the compressibility in this system as
is expected for FQH states.
As discussed in Ref.~\onlinecite{Nakamura-W-B},
a zero energy eigenstate can be created inserting $0$ everywhere in $\ket{\Psi^0_{1/q}}$.
This process makes a particle pair separated $q+1$ site
while the range of interactions is $q$ sites,
so that it is equivalent to making an open boundary.
Therefore the ground-state energy of such cases gives $E_q(N_s+1)=E_q(N_s)=0$.
On the other hand, if we shrink the system size by removing $0$,
a defect corresponding to the fractional charge $e^*=e/q$ appears.
Since this defect localizes,
the excitation energy $E_q(N_s-1)$ has very small size dependence as
was explicitly shown in the $\nu=1/3$ case.\cite{Nakamura-W-B}
Therefore, the inverse compressibility
\begin{equation}
\frac{1}{\kappa}=\lim_{N_s\to\infty}
\frac{N_s}{4\pi l_{\rm B}^2}
[E(N_s-1)+E(N_s+1)-2E_s(N_s)]
\end{equation}
clearly diverges.

\section{Matrix product method}\label{sec:MP}
From the exact solution (\ref{gwf1q}) it is possible to calculate
generic ground-state expectation values.  For this purpose, it is
convenient to introduce a matrix product (MP) representation
\cite{Fannes-N-W,Klumper-S-Z} of the ground-state wave function for
$\nu=1/q$ FQH states with $q=2m+1$ ($q=2m$) for fermions (bosons).  For
a periodic system with $N_s=q N$ sites, (\ref{gwf1q}) can be written as
\begin{equation}
 \ket{\Psi_{1/q}}= \tr[g_1 g_2 \cdots g_N],
  \label{MPS}
\end{equation}
where $g_i$ is identified as the following $(m+1)\times(m+1)$ matrix
(see Appendix \ref{sec:HowToGetMPS}),
\begin{equation}
 g_i=\left[
      \begin{array}{c c c c c}
       \ket{\rm o}_i &  \ket{+1}_i &  \ket{+2}_i &\cdots & \ket{+m}_i\\
       s_1\ket{-1}_i & 0 & 0 & \cdots & 0 \\
       s_2\ket{-2}_i & 0 & 0 & \cdots & 0 \\
       \vdots & \vdots & \vdots & \ddots & \vdots \\
       s_m\ket{-m}_i & 0 & 0 & \cdots & 0
      \end{array}
     \right].
\label{g-matrix2m+1}
\end{equation}
Here we have introduced the spin-$m$ representation for $q$-sites unit
cell for fermions (bosons),
\begin{equation}
\label{spin-m}
\begin{split}
&\ket{10\cdots 0\cdots 00}\; (\ket{00\cdots 00\cdots 00})\to\ket{-m},\\
&\ket{01\cdots 0\cdots 00}\; (\ket{10\cdots 00\cdots 00})\to\ket{-m+1},
\cdots,\\
&\ket{00\cdots 1\cdots 00}\; (\ket{00\cdots 10\cdots 00})\to\ket{\rm o},
\cdots,\\
&\ket{00\cdots 0\cdots 01}\; (\ket{00\cdots 00\cdots 02}) \to\ket{+m}.
\end{split}
\end{equation}
The matrix (\ref{g-matrix2m+1}) has the same size as that of the
valence-bond-solid states of an $S=m$ quantum spin
chain,\cite{Totsuka-S} however, there is a $m\times m$ zero entry part
and the transfer matrix $G(r_1,r_2;t_1,t_2)\equiv
g^{\dag}_j(r_1,t_1)g_j(r_2,t_2)$ has only two eigenvalues
$\lambda_{\pm}=(1\pm \sqrt{4\bm{s}^2-3})/2$.  Due to this property, the
matrix $G$ can be reduced to $4\times 4$ form for all $q$, so that the
calculation of expectation values is extremely simplified.

\begin{figure}[t]
 \includegraphics[width=85mm]{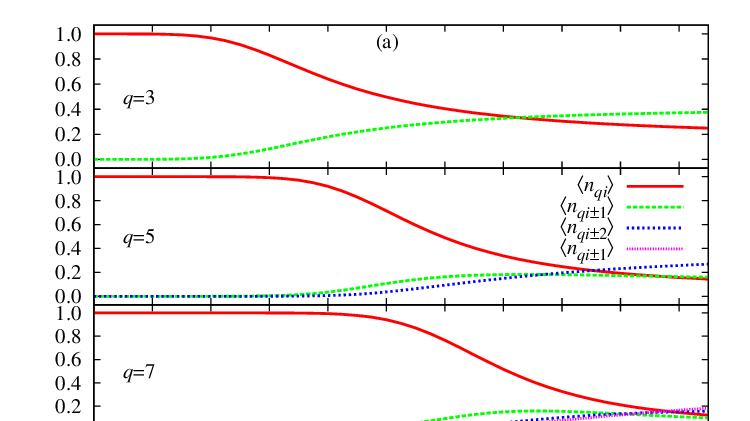}
 \includegraphics[width=85mm]{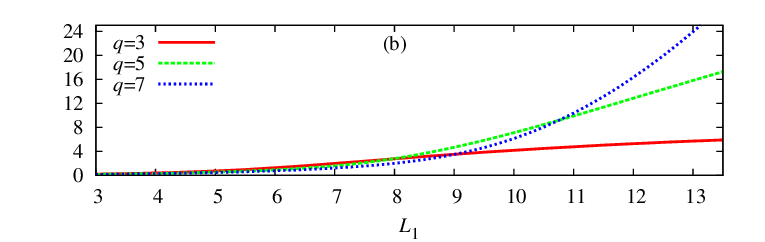}
 \caption{(Color online) (a) Density functions $\braket{\hat{n}_k}$ and
 (b) correlation lengths $\xi$ of the density-density correlation
 functions of $\nu=1/q$ FQH states with $q=3,5,7$ in $L_2\to\infty$
 limit as functions of the circumference of the torus $L_1$, calculated
 analytically by the MP method.}\label{fig2}
\end{figure}

\subsection{Correlation functions}
Using the MP formalism we obtain the density function in the infinite
systems $N \rightarrow \infty$, which has $q$-site periodicity as
(see Appendix \ref{sec:MPCF}),
\begin{align}
 \braket{\hat{n}_{qi \pm j}} 
 &=\frac{s_j^2}{2(\bm{s}^2-1)} 
  \left(
   1-\frac{1}{\sqrt{4\bm{s}^2-3}} \right),\\
  \braket{\hat{n}_{qi}}
 &=\frac{1}{\sqrt{4\bm{s}^2-3}}.
\end{align}
Figure \ref{fig2}(a) shows the $L_1$ dependence of the density functions
for $q=3,5,7$.  Since our wave function is not translationally
invariant, it does not describe a genuine liquid state, however, these
results indicate that the system's approach to homogeneous liquid states
as $L_1$ is increased. This is consistent with the result obtained by
Rezayi and Haldane which showed that the charge-ordered state in the TT
limit is adiabatically connected to the FQH liquid state through the
smooth deformation of the torus.\cite{Rezayi-H} Actually, our exact
result describes many important aspects of the FQH states, as we have
already seen.  Similarly, the density-density correlation functions are
calculated, which have exponentially decaying forms as (see Appendix
\ref{sec:MPCF})
\begin{equation}
 \braket{\hat{n}_{i}\hat{n}_{j}}
  \!-\!\braket{\hat{n}_{i}}\braket{\hat{n}_{j}}\!\sim\!
  \left(\!\frac{1-\sqrt{4\bm{s}^2-3}}{1+\sqrt{\bm{s}^2-3}\!}
  \right)^{|i-j|}\!\equiv \e^{-q|i-j|/\xi},
  \label{corr_func}
\end{equation}
where $\xi$ is the correlation length.  As shown in Fig.~\ref{fig2}(b),
the correlation length becomes longer in large $L_1$ regions as $q$ is
increased. Since the correlation length is proportional to the inverse
of the energy gap, this result is consistent with the fact that the
$\nu=1/q$ FQH state becomes unstable for larger $q$.  Similar behavior
has also been confirmed in the $\nu=p/(2p+1)$ Jain series.\cite{Wang-T-N}

\begin{figure}[t]
 \includegraphics[width=80mm]{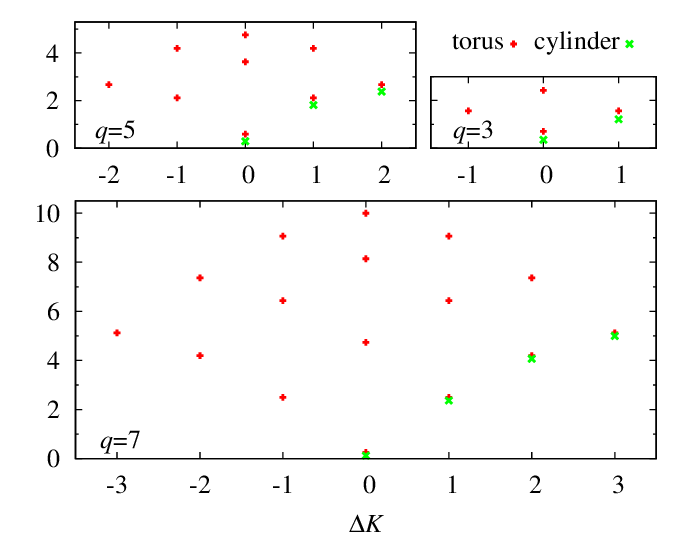}
 \caption{(Color online) The entanglement spectra of the fermion systems
 ($q=3,5,7$, $N \rightarrow \infty$) on torus (cylinder) geometries at
 $L_1=9$ as functions of the momentum for $x_1$ direction $\Delta K$.
 They form the conformal tower structure of (chiral) Tomonaga-Luttinger
 liquids describing the edge states.} \label{fig:ES}
\end{figure}

\begin{figure}[b]
 \includegraphics[width=80mm]{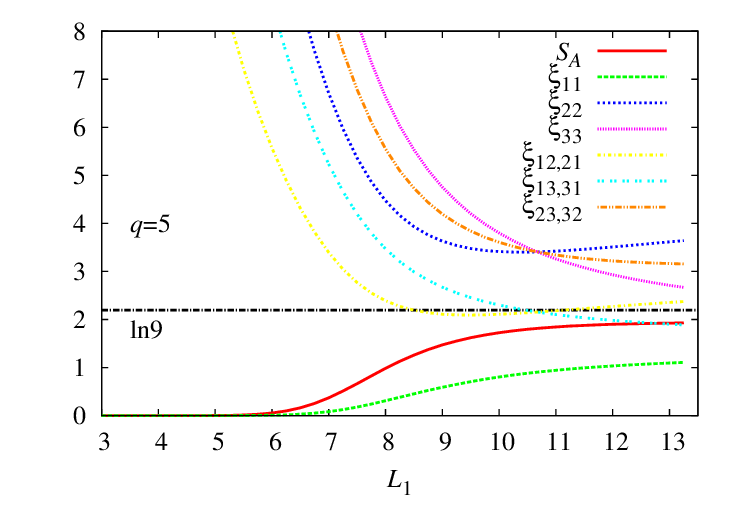}
 \caption{(Color online) The entanglement spectra $\xi_{ij}$ and entanglement
 entropy $S_A$ of the fermion systems ($q=5$, $N \rightarrow \infty$) on torus
 geometry as functions of the circumference of the torus $L_1$.}
 \label{fig:ESEE}
\end{figure}

\subsection{Entanglement spectra}
The MP formalism enables us to derive entanglement spectra (ES) of the
FQH system $\{\xi_{j} \}$ analytically (see Appendix
\ref{sec:MPES}). First we consider the cylinder geometry (open boundary
conditions), and divide the system into two blocks $A$ and $B$ with
length $L=N/2$. Then the wave function (\ref{gwf1q}) is written as the
Schmidt decomposition
\begin{equation}
 \ket{\Psi_{1/q}^{\mathrm{c}}}
  \equiv \sum_{j} \e^{-\xi_{j}/2}
  \ket{\psi^A_{1,j}}\otimes \ket{\psi^B_{j,1}},
\end{equation}
where $|\psi^{A(B)}_{j_1 j_2} \rangle = [\mathcal{N}^{A(B)}_{j_1
j_2}]^{-1/2} [\prod_{i=1}^{L} g_{i(+L)}]_{j_1 j_2}$ are orthogonal
states in the subsystem $A(B)$, and $\mathcal{N}^{A(B)}_{j_1 j_2}$ are
their norms. The ES are given by $\xi_j=-\ln(\mathcal{N}^A_{1,j}
\mathcal{N}^B_{j,1}/\mathcal{N})$.  As was pointed out by Li and
Haldane,\cite{Li-Haldane} the ES characterize the ``edge states'' of the
subsystem.  Although the number of the ES is limited as $m+1$, the
number of entanglement levels for bulk FQH states grows with the size of
the system.\cite{Li-Haldane} However, when the ES are plotted as
functions of the center-of-mass $\Delta K$ (which corresponds to
momentum of $x_1$ direction) of the subsystem as shown in
Fig.~\ref{fig:ES}, they give low-lying liner spectra of the gapless edge
state described as a chiral Tomonaga-Luttinger liquid.\cite{Wen}

Next, we evaluate ES in the torus geometry where the Schmidt
decomposition is given by
\begin{equation}
 | \Psi_{1/q}^{\mathrm{t}} \rangle \equiv
  \sum_{i j} \e^{-\xi_{ij}/2} |\psi^A_{i j} \rangle \otimes | \psi^B_{j i}
  \rangle.
\end{equation}
The ES $\xi_{ij}=-\ln(\mathcal{N}^A_{ij}
\mathcal{N}^B_{ji}/\mathcal{N})$ are characterized by two separated
``edge states'' of the subsystems.
The number of the ES is $(m+1)^2$, and $m(m+1)/2$ of them are two-fold
degenerate. As shown in Fig. \ref{fig:ES}, the ES show a ``diamond
shape'' conformal tower structure of a nonchiral Tomonaga-Luttinger
liquid which has both left and right movers.\cite{Lauchli-B-S-H} This
result can be explained by the relation obtained by the MP method
\begin{equation}
 \xi(\Delta K_1- \Delta K_2)=\xi(\Delta K_1)+ \xi(\Delta K_2)-\xi_0
  \label{ESNinf}
\end{equation}
for $N \rightarrow \infty$, where $\xi_0$ is the lowest ES. This means
that the ES are given by the spectra of an effective noninteracting 1D
``entanglement Hamiltonian'' with right and left modes with momenta
$\Delta K_1$, $-\Delta K_2$.
%
We can also evaluate the difference between ES on a torus and those of
cylinder as $\ln[\lambda_+/(\lambda_+ -\lambda_-)]$.

In Fig.~\ref{fig:ESEE}, $L_1$ dependence of the ES and entanglement
entropy (EE) for $q=5$ with torus boundary conditions are shown. The
diamond structure of ES is broken for the large $L_1$ regions and the EE
$S$ approach to $\ln(m+1)^2$.

\section{Conclusion} 
We have introduced a 1D minimal reference model with exact ground states
which describes $\nu=1/q$ FQH states with odd (even) $q$ for fermions
(bosons).  This model is constructed as the second quantization of the
pseudopotential for the Laughlin wave function, around the TT limit of
the torus boundary system.  The model has the same structures of the
pseudopotential, and naturally derives general properties of the
Laughlin wave function.  The ground states are described by the MP wave
functions that enable us to calculate physical quantities analytically.
Especially, we have analytically obtained the conformal tower structure
of the ES which reflects chiral Tomonaga-Luttinger liquid behavior of
the edge state.

\section{Acknowledgments}
We acknowledge support from the Global Center of Excellence Program
``Nanoscience and Quantum Physics'' of the Tokyo Institute of
Technology.  M.N. also acknowledges support from Grant-in-Aid
No.23540362 by MEXT.

\pagebreak

\appendix

\begin{widetext}

\section{Matrix elements of the pseudo potential [Eqs.~(\ref{TKint}) and
 (\ref{effTK})]}\label{sec:TKpotC}

We consider a model of $N$ interacting electrons on a torus with lengths
$L_l$ in $x_l$ directions ($l=1,2$).  When the torus is pierced by $N_s$
magnetic flux quanta, boundary conditions require $L_1L_2=2\pi N_s$. In
Landau gauge, $\bm{A}=Bx_2 \widehat{x}_1$, a complete basis of $N_s$
degenerate single-particle states in the lowest Landau level, labeled by
$k=0, \dots, N_s-1$, can be chosen as
\begin{equation}
\psi_k(\bm{x})=\frac{1}{\sqrt{\pi^{1/2} L_1}}
\sum_{m \in \mathbb{Z}}
\exp \left[ \mathrm{i}
\left(\frac{2\pi}{L_1}k+L_2m \right) x_1 \right]
\exp \left[ -\frac{1}{2}
\left(x_2
+\frac{2\pi }{L_1}k+L_2m \right)^2 \right],
\end{equation}
where unit of $x_l$ and $L_l$ is the magnetic length $l_{\rm B}$.  In
$L_2\to\infty$ limit, it becomes Eq.~(\ref{1particle_wf}).  In this
basis, any translation-invariant 2D Hamiltonian with two-body
interactions assumes the following 1D lattice model:
\begin{equation}
\begin{gathered}
\hat{\mathcal{H}}=\sum_{k_1k_2 k_3 k_4} \hat{V}_{k_1k_2k_3k_4}^{\mathstrut}
=\sum_{|m|\leq k \le N_s/2} \hat{V}_{km}; \\
\begin{split}
\hat{V}_{km}
&=\sum_{i} V_{km}^{\mathstrut}
c^\dagger_{i+m} c^\dagger_{i+k}
c_{i+k+m}^{\mathstrut}c_{i}^{\mathstrut}\\
&=\sum_{i} \hat{V}_{i+m,i+k,i+m+k,i} \mp \hat{V}_{i+m,i+k,i,i+m+k}
+\hat{V}_{i+k,i+m,i,i+m+k} \mp \hat{V}_{i+k,i+m,i+m+k,i},
\end{split}
\end{gathered}
\end{equation}
where the upper (lower) signs correspond to fermions (bosons) and the
operators $c_{k}^{\mathstrut}$($c_{k}^\dagger$) annihilate (create) a
particle at site (orbital) $k$.  The matrix elements
$\hat{V}_{q_1q_2q_3q_4}$ can be calculated according to the standard
second quantization
\begin{equation}
\hat{V}_{k_1k_2k_3k_4}=V_{k_1k_2k_3k_4}
c^\dagger_{k_1}c^\dagger_{k_2}c_{k_3}^{\mathstrut} c_{k_4}^{\mathstrut}
=\frac{1}{2} \iint \hat{\psi}^{\dagger}_{k_1}(\bm{x}_1)
\hat{\psi}^{\dagger}_{k_2}(\bm{x}_2)
V(\bm{x}_1-\bm{x}_2)\hat{\psi}^{\mathstrut}_{k_3}(\bm{x}_2)
\hat{\psi}^{\mathstrut}_{k_4}(\bm{x}_1)\mathrm{d}^2\bm{x}_1
 \mathrm{d}^2\bm{x}_2,
\end{equation}
where $\hat{\psi}^{\mathstrut}_{k}(\bm{x}) \equiv
\psi^{\mathstrut}_{k}(\bm{x}) c_k^{\mathstrut}$.  One can find that for
fermionic (bosonic) systems the matrix elements are real and
antisymmetric (symmetric) functions for $k$ and $m$ reflecting Fermi
(Bose) statistics.  Using the Fourier transform $V(\bm{p})$ as the
interaction potential $V(\bm{x}_1-\bm{x}_2)$, the matrix elements
$V_{k_1k_2k_3k_4}$ are given by
\begin{equation}
 V_{k_1k_2k_3k_4}=
  \frac{\delta'_{k_1+k_2 ,k_3+k_4}}{2L_1L_2}
  \sum_{\bm{p}} \delta'_{k_1-k_4, p_1L_1/2\pi}
  V(\bm{p}) \mathrm{e}^{-\bm{p}^2/2-\mathrm{i}(k_1-k_3)p_2L_2/N_s},
\end{equation}
where $\bm{p}={}^t(p_1, p_2)$, $p_1=2\pi m_1 /L_1$, $p_2=2\pi m_2 /L_2$,
$m_1,m_2 \in \mathbb{Z}$, and $\delta'$ is the Kronecker delta function
with a period $N_s$.  The pseudo potential proposed by Trugman and
Kivelson is known that it has the $\nu=1/q$ Laughlin wave function as
the exact ground state,\cite{Trugman-K}
\begin{equation}
 \mathcal{V}(\bm{x})=\sum_{\lambda=0}^{q-1}\mathcal{V}^{(\lambda)}
=\sum_{\lambda=0}^{q-1}c_{\lambda}b^{2\lambda}
  \nabla^{2\lambda}\delta^2(\bm{x}), \quad
\bm{x}=\bm{x}_1-\bm{x}_2,
\end{equation}
where $b$ is the range of the interaction. 
For $N_s \gg L_1$, the matrix elements $\mathcal{V}_{km}$ are calculated
in the $L_2\to\infty$ limit using the binomial theorem and Gauss
integrals $S_n\equiv\int_{-\infty}^{\infty}dx\, x^{2n}\e^{-x^2/2}$ as
\begin{equation}
\begin{split}
 \mathcal{V}_{km}^{(\lambda)}
 =&\frac{(- b^2)^{\lambda}c_{\lambda}}{4 \pi L_1}
 \mathrm{e}^{-( K^2+M^2)/2}\nonumber\\
 &\times\int \mathrm{d}x
 \left[ 
 (M^2+x^2)^\lambda
 \left(
 \mathrm{e}^{-(x-\mathrm{i}K)^2/2}
 +\mathrm{e}^{-(x+\mathrm{i}K)^2/2}
 \right)
 \mp (K^2+x^2)^\lambda
 \left(
 \mathrm{e}^{- (x-\mathrm{i}M)^2/2}
 +\mathrm{e}^{-(x+\mathrm{i}M)^2/2}
 \right)
 \right]\\
 =&\frac{(- b^2)^{\lambda}c_{\lambda}}{4 \pi L_1}
 \mathrm{e}^{-( K^2+M^2 )/2}
 \sum_{\nu=0}^{\lambda}\ _{\lambda}\mathrm{C}_{\nu}
 ( K^2 -M^2 )^{\nu}\\
 &\times \int \mathrm{d}x
 \left[
 (-1)^{\nu}( K^2 +x^2)^{\lambda-\nu}
 \left(
 \mathrm{e}^{-(x-\mathrm{i}K)^2/2}
 +\mathrm{e}^{-(x+\mathrm{i}K)^2/2}
 \right)
 \mp ( M^2 +x^2)^{\lambda-\nu}
 \left(
 \mathrm{e}^{-(x-\mathrm{i}M)^2/2}
 +\mathrm{e}^{-(x+\mathrm{i}M)^2/2}
 \right)
 \right] \\
 =&\frac{(- b^2)^{\lambda}c_{\lambda}}{4 \pi L_1}
 \mathrm{e}^{-(K^2+M^2)/2}
\sum_{\nu=0}^{\lambda}
\ _{\lambda}\mathrm{C}_{\nu}
 (K^2 -M^2)^{\nu}
 \sum_{\mu=0}^{[(\lambda-\nu)/2]}
 (-1)^{\mu}
 2^{2\mu+1}\ _{\lambda - \nu}\mathrm{C}_{2\mu}
S_{\lambda - \nu-\mu}
 \left[ (-1)^{\nu}K^{2\mu}\mp M^{2\mu}\right] \\
 =&\frac{(- b^2)^{\lambda}c_{\lambda}}{2 \sqrt{2\pi} L_1}
 \mathrm{e}^{-( K^2+M^2)/2}
\sum_{\nu=0}^{\lambda}
 \sum_{\mu=0}^{[(\lambda-\nu)/2]}
 (-1)^{\mu}
 2^{2\mu+1}
 \frac{\lambda ! (2\lambda-2\nu-2\mu-1)!!}
 {\nu! (2\mu)!  (\lambda-\nu-2\mu)!}
 ( K^2 -M^2 )^{\nu}
 \left[ (-1)^{\nu}K^{2\mu}\mp M^{2\mu}\right]
\end{split}
\end{equation}
where $K=\frac{2 \pi k}{L_1}$ and $M=\frac{2 \pi m}{L_1}$.  One can find
that in fermion (boson) systems the powers of $(K^2-M^2)$ are always odd
(even) for each term, and the leading term of $K$, $M$ appears as
$(K^2-M^2)^{\lambda}$ when $\lambda$ is odd (even), because
$\left[(-1)^{\nu}K^{2\mu}\mp M^{2\mu}\right]$ vanishes for
$(\nu,\mu)=(\lambda,0)$ when $\lambda$ is even (odd). Therefore
Eq.~(\ref{TKint}) can be approximated as Eq.~(\ref{effTK}) in the small
$L_1$ regions.

\end{widetext}

\section{Matrix Product Representation [Eq.~(9)]}

\label{sec:HowToGetMPS}
Let us consider the case of $\nu=1/3$ as the simplest example.  The
ground-state manifold of this model is threefold degenerate, 
spanned by charge ordered states with one electron per a three-site unit cell:
$\ket{\cdots\ 010\ 010\ 010\ \cdots}$.  The $\hat V_{21}$ term induces
fluctuations upon these ground states through the process
\begin{equation}
 \ket{010\ 010}\leftrightarrow\ket{001\ 100}.
\end{equation}
The truncated model can be mapped to an $S=1$ quantum spin chain by
identifying the states of the unit cell as
$\ket{010}\to\ket{\mathrm{o}}$, $\ket{001}\to\ket{+}$, and
$\ket{100}\to\ket{-}$.  We consider the matrix product representation of
the ground-state wave function in the $S=1$ spin basis.
Then the matrix can be written in the following $3 \times 3$ form
\begin{equation}
\label{f-matrix}
g_i=\left[
\begin{array}{c c c}
 f_{--}\ket{- }_i  &  f_{-\mathrm{o}}\ket{\mathrm{o} }_i &   f_{-+} \ket{+ }_i \\
 f_{\mathrm{o}-}\ket{- }_i  & f_{\mathrm{oo}}\ket{\mathrm{o} }_i & f_{\mathrm{o}+} \ket{+ }_i \\
 f_{+-}\ket{- }_i  &  f_{+\mathrm{o}}\ket{\mathrm{o} }_i &  f_{++}\ket{+ }_i
\end{array}
\right],
\end{equation}
where $i$ is the site index for the unit cell and $f_{\mathrm{IO}}$ are
constants.  The index $\mathrm{I}$ and $\mathrm{O}$ denote the input and
output states that are the configurations of the $(i-1)$th and the
$i$th spins.  We suppose that the ground-state wave function in
periodic boundary conditions is given by
\begin{align}
\ket{\Psi_{1/3} }_\mathrm{p}=&
 \prod_{i=1}^{N}(1 + s_1 \hat{U}_{3i-1}) \ket{0100100100...},\\
=& \prod_{i=1}^{N} (1 + s_1 \hat{\mathcal{U}}_i) \ket{\mathrm{ooo}...}\\
=&\mathrm{tr}[g_1 g_2 \cdots g_N],
\end{align}
where $s_1 \equiv\alpha_0/\alpha_1$, and
\begin{align}
 \hat{U}_{k}^{\mathstrut}=c^\dagger_{k+1}c^\dagger_{k+2}
 c_{k+3}^{\mathstrut} c_{k}^{\mathstrut},\quad
 \hat{\mathcal{U}}_i=
-\frac{1}{2} \hat{S}_i^z\hat{S}_i^+ \hat{S}_{i+1}^z\hat{S}_{i+1}^-,
\end{align}
with $\hat{S}^{\alpha}_i$ being an $S=1$ spin operator.
One can find that the states of the nearest two spins can be written as
\begin{equation}
\psi_{i-1} \otimes \psi_i = \ket{\mathrm{oo}} + s_1 \ket{\mathrm{+}-} + \ket{\mathrm{o}+}
+\ket{-\mathrm{o}}+ \ket{-+}.
\end{equation}
Therefore, we can set
\begin{align}
& f_{--}=f_{\mathrm{o}-}= f_{+\mathrm{o}}=f_{++}=0 \\
& f_{\mathrm{oo}}=f_{\mathrm{o}+}= f_{-\mathrm{o}}= f_{-+}= 1 \\
& f_{+-}=s_1.
\end{align}
Thus, the matrix Eq.~(\ref{f-matrix}) then becomes
\begin{equation}
g_i=\left[
\begin{array}{c c c}
0 & \ket{\mathrm{o} }_i &  \ket{+ }_i \\
0 & \ket{\mathrm{o} }_i &  \ket{+ }_i \\
s_1\ket{- }_i  & 0 &  0
\end{array}
\right].
\label{g-matrix}
\end{equation}
If we change the matrix (\ref{f-matrix}) in different basis
$(s_1\ket{-}_i+\ket{\mathrm{o}}_i,\ket{+}_i)$, the above matrix can be
reduced to $2\times 2$ form without losing generality,\cite{Nakamura-W-B}
\begin{equation}
g_i=\left[
\begin{array}{c c}
 \ket{\mathrm{o}}_i &  \ket{+}_i \\
 s_1\ket{-}_i  & 0
\end{array}
\right].\label{g-reduced_}
\end{equation}
Although Eqs.~(\ref{g-matrix}) and (\ref{g-reduced_}) give the same
result, we use the latter representation for simplicity.

In the diagonal component of the matrix product, the upper one is both
for open and periodic boundary conditions, while the lower one is only
for periodic boundary systems.  Therefore, for open boundary systems,
the ground-state wave function is given as
\begin{equation}
 \ket{\Psi_{1/3}}_\mathrm{o}=\mathrm{tr}'[g_1 g_2 \cdots g_N],
\end{equation}
where $\mathrm{tr}'$ means partial trace for the upper component.

For general $\nu=1/q$ with $q=2m+1$, we consider in a similar way as the
above using the spin-$m$ representation (\ref{spin-m}), and obtain a $q
\times q$ matrix as
\begin{equation}
g_i=
 \left[
  \begin{array}{c c c c c c c}
   0  &  \cdots & 0 & \ket{\mathrm{o}} & \ket{+1} & \cdots & \ket{+m} \\
   \vdots & \ddots & \vdots & \vdots & \vdots & \vdots & \vdots   \\
   0 & \cdots & 0 & \ket{\mathrm{o}} & \ket{+1} & \cdots & \ket{+m} \\
   \vdots & \iddots & \ket{-1} & 0 &  \cdots & \cdots & 0 \\
   0 & \iddots  & \iddots & \vdots &  \cdots & \cdots & 0 \\
   \ket{-m} & 0  & \cdots & 0 &  \cdots & \cdots & 0 \\
  \end{array}
 \right].
\end{equation}
The matrix with the reduced dimension $m+1$ can be obtained as
Eq.~(\ref{g-matrix2m+1}) in the text.

\section{Correlation functions}

\subsection{Transfer matrix}
We consider the matrix product representation of the $\nu=1/(2m+1)$
ground-state wave function in $(2m+1)N$ sites periodic system.  The
normalized ground-state wave function can be written as
\begin{align}
  \label{MPS}
 \ket{\Psi_m}
 &={\cal N}^{-1/2}\,\mathrm{Tr} [g_1 g_2 \cdots g_N],\\
 \mathcal{N} &\equiv \mathrm{Tr} \left[ G^N \right], \qquad
G \equiv \bar{g}_{i} \otimes g_{i},
\end{align}
where $g_i$ is identified as Eq.~(\ref{g-matrix2m+1}).  The $(m+1)^2
\times (m+1)^2$ transfer matrix $G(r_1,r_2;t_1,t_2)\equiv
g^{\dag}_j(r_1,t_1)g_j(r_2,t_2)$ has only two eigenvalues,
\begin{equation}
 \lambda_0=0,\qquad
 \lambda_{\pm}
=\frac{1\pm \sqrt{4\bm{s}^2-3}}{2},
\end{equation}
that are obtained from the following secular equation:
\begin{equation}
0=\det \left| G- \lambda \right| =(1-\lambda)(-\lambda)^m
-(-\lambda)^{m-1}\sum_{j=1}^m s^2_j.
\end{equation}
We define $(m+2) \times (m+2)$ local matrix $\tilde{1}$
to reduce the dimension of $G$ to $(m+1) \times (m+1)$:
\begin{align}
G=\left[
      \begin{array}{c c c c}
       \tilde{1} & \tilde{1} & \cdots & \tilde{1} \\
       (s_1^2\tilde{1}) &  0 & \cdots & 0 \\
        \vdots &  \vdots & \ddots & \vdots \\
      (s_m^2\tilde{1}) & 0 & \cdots & 0 \\
      \end{array}
     \right],\quad
\tilde{1} =\left[
      \begin{array}{c c c c}
       1 &  0 & \cdots & 0 \\
       0 &  0 & \ddots & \vdots \\
       \vdots & \ddots & \ddots & 0 \\
       0 & \cdots & 0 & 0 \\
      \end{array}
     \right].
\end{align}
Thus the transfer matrix is diagonalized by finding a unitary matrix $U$
as
\begin{equation}
G_0=U^{-1}G U=\left[
      \begin{array}{c c c c c }
       \left(\lambda_- \tilde{1}\right) & 0 & \cdots & \cdots & 0 \\
       0 & \left(\lambda_+ \tilde{1}\right) & 0 & \cdots & 0 \\
       \vdots & 0 & 0 & \cdots & 0 \\
       \vdots & \vdots & \vdots & \ddots & \vdots \\
       0 & 0 & 0 & \cdots & 0 \\
      \end{array}
     \right].
\label{DefU}
\end{equation}
Thus we can calculate physical quantities in just the same way as in the
case of $\nu=1/3$.\cite{Nakamura-W-B}

\subsection{Density and Charge correlation}
\label{sec:MPCF}
In the spin-$m$ representation for the $\nu=1/(2m+1)$ state
(\ref{spin-m}), the density operator and its transfer matrix can be
written as
\begin{equation}
 \hat{n}_{qi + j}=\delta_{\hat{S}_i^z, j},\qquad
  G[\hat{n}](j)\equiv \bar{g}_i \otimes \delta_{\hat{S}_i^z,j} g_i \, .
\end{equation}
Using the unitary transformation $U$ defined in Eq.~(\ref{DefU}), we
obtain
the density function with $q$ sites periodicity as
\begin{equation}
\begin{split}
\mathcal{D}(j ; N) &\equiv \left\langle \hat{n}_{qi + j} \right\rangle
=\left\langle \hat{n}_{qi - j} \right\rangle\\
&=\frac{\mathrm{Tr}\left[ U^{-1} G[\hat{n}](|j|) U G_0^{N-1} \right]}
{\mathrm{Tr}\left[ G_0^{N} \right]},\\
\lim_{N \to \infty} \mathcal{D}(j ; N)&=
\left\{
\begin{array}{ l l }
\frac{s_j^2}{2(\bm{s}^2-1)}\left( 1-\frac{1}{\sqrt{4\bm{s}^2-3}} \right) & (j \neq 0) \\
\frac{1}{\sqrt{4\bm{s}^2-3}} & (j = 0) \\
\end{array}
\right. \, .
\end{split}
\end{equation}
The charge-charge correlation function is calculated as
\begin{equation}
\begin{split}
&\mathcal{C}(i, j ,i', j'; N) 
\equiv \left\langle \hat{n}_{qi + j}  \hat{n}_{qi' + j'} \right\rangle\\
&=\frac{\mathrm{Tr}\left[ U^{-1} G[\hat{n}](j) U G_0^{(i'-i-1)}
U^{-1} G[\hat{n}](j') U G_0^{(N-1-i'+i)}
\right]}
{ \mathrm{Tr}\left[ G_0^{N} \right]},
\end{split}
\end{equation}
\begin{align}
\lefteqn{\lim_{N,|i-i'| \to \infty} \mathcal{C}(i, j ,i', j'; N) \sim 
  \left(\frac{\lambda_-}{\lambda_+}
  \right)^{|i-i'|}}\nonumber\\
&=\left( \frac{1-\sqrt{4\bm{s}^2-3}}{1+\sqrt{4\bm{s}^2-3}} \right)^{|i-i'|}
\equiv \e^{-q|i-i'|/\xi} ,
\end{align}
where $\xi$ is the correlation length for the infinite system.

\section{Entanglement Spectra and Entropy}\label{sec:MPES}
In this section, we compute the entanglement spectrum (ES) and the
entanglement entropy (EE) both in cylinder and torus geometries. We
obtain the conformal tower structure of the ES in Fig.~\ref{fig:ES}
analytically.

\subsection{Cylinder}
For a open boundary system, the normalized ground-state wave function
can be written as:
\begin{equation}
 \ket{\Psi_m}={\cal N}^{-1/2}\,\mathrm{Tr} [g'_1 g_2 \cdots g''_N],  \quad
\mathcal{N} \equiv \mathrm{Tr} \left[ G' G^{N-2} G'' \right],
  \label{MPSop}
\end{equation}
where the terminators $g'_i , g''_i$ are identified as the following
matrices
\begin{align}
g'_i &\equiv \left[
      \begin{array}{c c c c }
       \ket{ 0}_i &  \ket{+1}_i &  \cdots & \ket{+m}_i\\
       0 & 0 & \cdots & 0 \\
       \vdots & \vdots & \ddots & \vdots \\
        0 & 0 & \cdots & 0
      \end{array}
     \right],\\
g''_i &\equiv \left[
      \begin{array}{c c c c }
       \ket{ 0}_i &  0 &  \cdots & 0\\
       s_1\ket{-1}_i & 0  & \cdots & 0 \\
       \vdots & \vdots & \ddots & \vdots \\
        s_m\ket{-m}_i & 0 & \cdots & 0
      \end{array}
     \right] , 
\end{align}
and the transfer matrices $G', G''$ are given by
\begin{align}
G' \equiv \bar{g}'_i \otimes g'_i &=
\left[
      \begin{array}{c c c c}
       \tilde{1} & \tilde{1} & \cdots & \tilde{1} \\
      0 &  0 & \cdots & 0 \\
        \vdots &  \vdots & \ddots & \vdots \\
      0 & 0 & \cdots & 0 \\
      \end{array}
     \right] ,\\
G'' \equiv \bar{g}''_i \otimes g''_i &=
\left[
      \begin{array}{c c c c}
       \tilde{1} & 0 & \cdots & 0 \\
       (s_1^2\tilde{1}) &  0 & \cdots & 0 \\
        \vdots &  \vdots & \ddots & \vdots \\
      (s_m^2\tilde{1}) & 0 & \cdots & 0 \\
      \end{array}
     \right].
\end{align}
We cut this system into blocks $A$ and $B$ with length $L$ and $N-L$,
respectively.  The wave function of the whole system is divided as
\begin{align}
\lefteqn{
| \Psi_m \rangle =\mathcal{N}^{-1/2}
\mathrm{Tr} \left[ g'_{1}g_{2}\cdots g''_{N} \right]} \\
&=\mathcal{N}^{-1/2}\sum_{j=1}^{m+1} \left\{  
\left[ g'_{1}g_{2}\cdots g_{L} \right]_{1, j} 
\left[ g_{L+1}g_{L+2}\cdots g''_N \right]_{j, 1}
\right\}.
\end{align}
We can then define the block states as
\begin{align}
| \psi^A_{j} \rangle&\equiv(\mathcal{N}^A_{j})^{-1/2} 
\big[ g'_1 g_2 \cdots g_L \big]_{1, j},\\
| \psi^B_{j} \rangle&\equiv(\mathcal{N}^B_{j})^{-1/2} 
\big[ g_{L+1} g_{L+2} \cdots g''_N \big]_{j, 1} ,
\end{align}
where $\mathcal{N}^{A(B)}_{j}$ is the normalization factors that is
given by
\begin{align}
\label{DefNop}
\mathcal{N}^A_{j}&\equiv\left[ G' G^{L-1} \right]_{j(m+2)-(m+1),1},\\
\mathcal{N}^B_{j}&\equiv\left[ G^{N-L-2} G'' \right]_{1,j(m+2)-(m+1)}.
\end{align}
The reduced density matrix is given by
\begin{equation}
\hat{\rho}^A = \sum_{j}
\frac{\mathcal{N}^A_{j} \mathcal{N}^B_{j}}{\mathcal{N}} \,
| \psi^A_{j} \rangle \langle \psi^A_{j} | .
\end{equation}
The ES $\xi^A_{j}$ and EE $S^A \equiv-\mathrm{Tr}_A \left[ \hat{\rho}_A
\log \hat{\rho}_A \right]$ can be written as
\begin{align}
\label{DefXiSop}
\xi^A_{j}&=-\log \left(
 \frac{\mathcal{N}^A_{j} \mathcal{N}^B_{j}}{\mathcal{N}}
\right),\\
S^A
&= -\sum_{j} \frac{\mathcal{N}^A_{j} \mathcal{N}^B_{j}}{\mathcal{N}}
\log \left(  \frac{\mathcal{N}^A_{j} \mathcal{N}^B_{j}}{\mathcal{N}} \right) .
\end{align}
The normalization factor for the wave function is calculated by using
Eq.~(\ref{MPSop}) as
\begin{equation}
\mathcal{N}
=\frac{\lambda_-^{N+1}-\lambda_+^{N+1}}{\lambda_--\lambda_+}.
\end{equation}
For $L=N/2 \Rightarrow G^L = G^{N-L}= G^{N/2}$, the normalization
factors $\mathcal{N}^A_{j}, \mathcal{N}^B_{j}$ are written as
\begin{align}
\mathcal{N}^A_{j}&=
\left\{
\begin{array}{l l}
\frac{\lambda_-^{L+1} - \lambda_+^{L+1}}{\lambda_--\lambda_+} &
(j =1) \\
\frac{\lambda_-^{L} - \lambda_+^{L}}{\lambda_--\lambda_+} &
(j >1) \\
\end{array}
\right.,\\
\mathcal{N}^B_{ j}&=
\left\{
\begin{array}{l l}
\frac{\lambda_-^{L+1} - \lambda_+^{L+1}}{\lambda_--\lambda_+} &
(j=1) \\
\frac{\lambda_-^{L} - \lambda_+^{L}}{\lambda_--\lambda_+}s_{j-1}^2 &
(j>1) \\
\end{array}
\right. .
\end{align}
We now can obtain the ES and EE by substituting the $\mathcal{N}^A_{j},
\mathcal{N}^B_{j}$ into Eq.~(\ref{DefXiSop}).  For $L=N/2 \rightarrow
\infty$,
\begin{equation}
\frac{\mathcal{N}^A_{j}\mathcal{N}^B_{j} }{\mathcal{N}} \rightarrow
\left\{
\begin{array}{l l}
\frac{ - \lambda_+}{\lambda_--\lambda_+} &
(j =1) \\
\frac{-1}{(\lambda_--\lambda_+)\lambda_+}s_{j-1}^2 &
(j>1) \\
\end{array}
\right. .
\end{equation}
When the ES are plotted as functions of the shift of the center of mass
$\Delta K$ of the subsystem, it shows approximately a liner relation
which corresponds to a conformal tower of the chiral Tomonaga-Luttinger
liquid.

\subsection{Torus}
We cut a periodic system into blocks $A$ and $B$ with lengths $L$ and $N-L$,
respectively.  The wave function of the whole system (\ref{MPS}) can be
divided as
\begin{align}
\lefteqn{\Psi_m \rangle=\mathcal{N}^{-1/2}
 \mathrm{Tr} \left[ g_{1}g_{2}\cdots g_{N} \right]} \\
&=\mathcal{N}^{-1/2}\sum_{j_1, j_2=1}^{m+1} 
 \left\{  
 \left[ g_{1}g_{2}\cdots g_{L} \right]_{j_1 j_2} 
 \left[ g_{L+1}g_{L+2}\cdots g_N \right]_{j_2 j_1}
 \right\}.
\end{align}
We can then define the block states as
\begin{align}
| \psi^A_{j_1 j_2} \rangle
 &\equiv(\mathcal{N}^A_{j_1 j_2})^{-1/2} 
\big[ g_1 g_2 \cdots g_L \big]_{j_1 j_2},\\
| \psi^B_{j_1 j_2} \rangle
 &\equiv(\mathcal{N}^B_{j_1 j_2})^{-1/2} 
\big[ g_{L+1} g_{L+2} \cdots g_N \big]_{j_2 j_1},
\end{align}
where $\mathcal{N}^{A(B)}_{j_1 j_2}$ is the normalization factors for
A(B) block,
\begin{align}
\label{DefN}
 \mathcal{N}^A_{j_1 j_2}
 &\equiv\left[ G^L \right]_{j_1(m+2)-(m+1),j_2(m+2)-(m+1)},\\
 \mathcal{N}^B_{j_1 j_2}
 &
 \equiv\left[ G^{N-L} \right]_{j_2(m+2)-(m+1),j_1(m+2)-(m+1)}.
\end{align}
One can now rewrite the ground-state wave function as Schmidt
decomposition,
\begin{equation}
| \Psi_m \rangle = \sum_{j_1 j_2} 
\left(
\frac{\mathcal{N}^A_{j_1 j_2} \mathcal{N}^B_{j_1 j_2}}{\mathcal{N}} 
\right)^{1/2}
| \psi^A_{j_1 j_2} \rangle \otimes | \psi^B_{j_1 j_2}  \rangle .
\end{equation}
The reduced density matrix is given by
\begin{equation}
\hat{\rho}^A = \sum_{j_1 j_2}
\frac{\mathcal{N}^A_{j_1 j_2} \mathcal{N}^B_{j_1 j_2}}{\mathcal{N}} \,
| \psi^A_{j_1 j_2} \rangle \langle \psi^A_{j_1 j_2} | .
\end{equation}
The ES $\xi^A_{j_1 j_2}$ and EE $S^A$ can be written as
\begin{align}
\label{DefXiS}
\xi^A_{j_1 j_2}&=-\ln \left(
 \frac{\mathcal{N}^A_{j_1 j_2} \mathcal{N}^B_{j_1 j_2}}{\mathcal{N}}
\right),\\
S^A &= \sum_{j_1 j_2} \frac{\mathcal{N}^A_{j_1 j_2} 
 \mathcal{N}^B_{j_1 j_2}}{\mathcal{N}}
 \ln \left(  \frac{\mathcal{N}^A_{j_1 j_2} 
\mathcal{N}^B_{j_1 j_2}}{\mathcal{N}} \right).
\end{align}
For $L=N/2$, the normalization factors $N_{j_1j_2}^A$,$N_{j_1j_2}^B$ are
calculated as
\begin{align}
\mathcal{N}^A_{j_1 j_2}&=
\left\{
\begin{array}{l l}
\frac{\lambda_-^{L+1} - \lambda_+^{L+1}}{\lambda_--\lambda_+} &
(j_1=1 ,\, j_2 =1) \\
\frac{\lambda_-^{L} - \lambda_+^{L}}{\lambda_--\lambda_+}s_{j_1-1}^2 &
(j_1>1,\, j_2 =1) \\
\frac{\lambda_-^{L} - \lambda_+^{L}}{\lambda_--\lambda_+} &
(j_1=1,\, j_2 >1) \\
\frac{\lambda_-^{L-1} - \lambda_+^{L-1}}{\lambda_--\lambda_+}s_{j_1-1}^2 &
(j_1>1,\, j_2 >1) \\
\end{array}
\right.,\\
\mathcal{N}^B_{j_1 j_2}&=\mathcal{N}^A_{j_2 j_1}.
\end{align}
Now we can use Eq.~(\ref{DefXiS}) to calculate the ES and the EE.  For
$L=N/2 \rightarrow \infty$, we get
\begin{align}
\frac{\mathcal{N}^A_{j_1 j_2}}{\sqrt{\mathcal{N}}} &\rightarrow
\left\{
\begin{array}{l l}
\frac{ - \lambda_+}{\lambda_--\lambda_+} &
(j_1=1 ,\, j_2 =1) \\
\frac{-1}{\lambda_--\lambda_+}s_{j_1-1}^2 &
(j_1>1,\, j_2 =1) \\
\frac{-1}{\lambda_--\lambda_+} &
(j_1=1,\, j_2 >1) \\
\frac{ - \lambda_+^{-1}}{\lambda_--\lambda_+}s_{j_1-1}^2 &
(j_1>1,\, j_2 >1) \\
\end{array}
\right.,\\
\mathcal{N}^B_{j_1 j_2} &=\mathcal{N}^A_{j_2 j_1}.
\end{align}
When the ES are plotted as functions of the shift of the center-of-mass
$\Delta K$ of the subsystem, for $N \rightarrow \infty$, we can show the
relation (\ref{ESNinf}) which means that the ES of the torus are given
by the combination of the right and the left movers of the chiral
Tomonaga-Luttinger liquids. Therefore, the ES shows the conformal tower
structure of the nonchiral Tomonaga-Luttinger liquid.  We can estimate
the difference of the ES on a cylinder and that on a torus as $\ln
[\lambda_+/(\lambda_+ -\lambda_-)]$.

\end{document}